\documentclass[prd,aps,fleqn,11pt]{revtex4} 
\usepackage[figuresright]{rotating}  
\usepackage{amsfonts}
\usepackage{amssymb,amsmath}

\newcommand{\et}{{\it et al.}}
\newcommand{\refcite}[1]{\cite{#1}}

\begin{document}


\title{CHARMONIUM SPECTROSCOPY FROM LATTICE QCD\footnote{ To appear in
    the Proceedings of the Beijing CHARM2010 Conference, October
    21-24, 2010.}}

\author{CARLETON DeTAR}

\affiliation{Department of Physics and Astronomy, University of Utah,
Salt Lake City, UT 84112, USA\footnote{
Temporary address: University of Tsukuba, Tsukuba, Japan}}

\begin{abstract}
I give a short review of the current status of lattice QCD calculations
of the spectrum of charmonium.
\end{abstract}


\maketitle

\section{Introduction}

In recent years the numerical simulation of lattice quantum
chromodynamics (QCD) has developed into an enormously successful
method for obtaining precise values of several Standard Model
parameters.  It is the only truly {\it ab initio} method available for
doing nonperturbative studies of QCD.  It provides a unified framework
for treating mesons and baryons involving charm and bottom quarks,
whether in a heavy-light or heavy-heavy configuration.

For charm physics the primary phenomenological objectives of lattice
QCD are to help in the discovery and characterization of excited
states, including exotics, to determine decay
constants\cite{Simone,Na} and form factors, and to calculate
electromagnetic transition rates.  Theoretical objectives include
providing guidance for effective field theory.

Since the last Charm conference there have been no new comprehensive
numerical lattice studies of exotics and excited quarkonium states.
So I will review the current status only briefly.  There has been
exciting progress, however, in the high precision determination of
masses of several of the lowest lying states involving heavy quarks.
Progress here is, of course, prerequisite to an accurate determination
and characterization of excited states.

Lattice simulations have their limitations.  While static properties,
including masses and matrix elements are relatively easy to calculate,
scattering processes, including charmonium production, inclusive
processes, and multihadronic (more than two-body) decays are very
difficult.

In addition to the usual challenges of extrapolating to the continuum
and physical light quark masses, heavy quark simulations face special
difficulties.  The lattice spacing $a$ introduces a momentum cut off
of order $1/a$, typically less than $2-3$ GeV with today's lattices.
As the quark mass $M$ approaches the cut off, discretization errors
grow.  Such mass-related errors can be controlled by a suitable
formulation of the Dirac action for the heavy quark.  Among them are
(1) a lattice version of nonrelativistic QCD\cite{Lepage:1992tx},
which converges rather slowly for charm, but well for bottom, (2) the
Fermilab action\cite{EKM} and its
improvements\cite{Oktay:2008ex,Oktay:LAT2010}, good for both charm and
bottom, and (3) the highly improved staggered quark action (HISQ) with
discretization errors of order ${\cal O}(\alpha_s^2(a M_c)^2)$, good
for charm, but not so good for bottom with today's lattices.

\begin{figure}
  \begin{tabular}{cc}
    \begin{minipage}{0.45\textwidth}
      \vspace*{-3mm}
      \includegraphics[width=\textwidth]{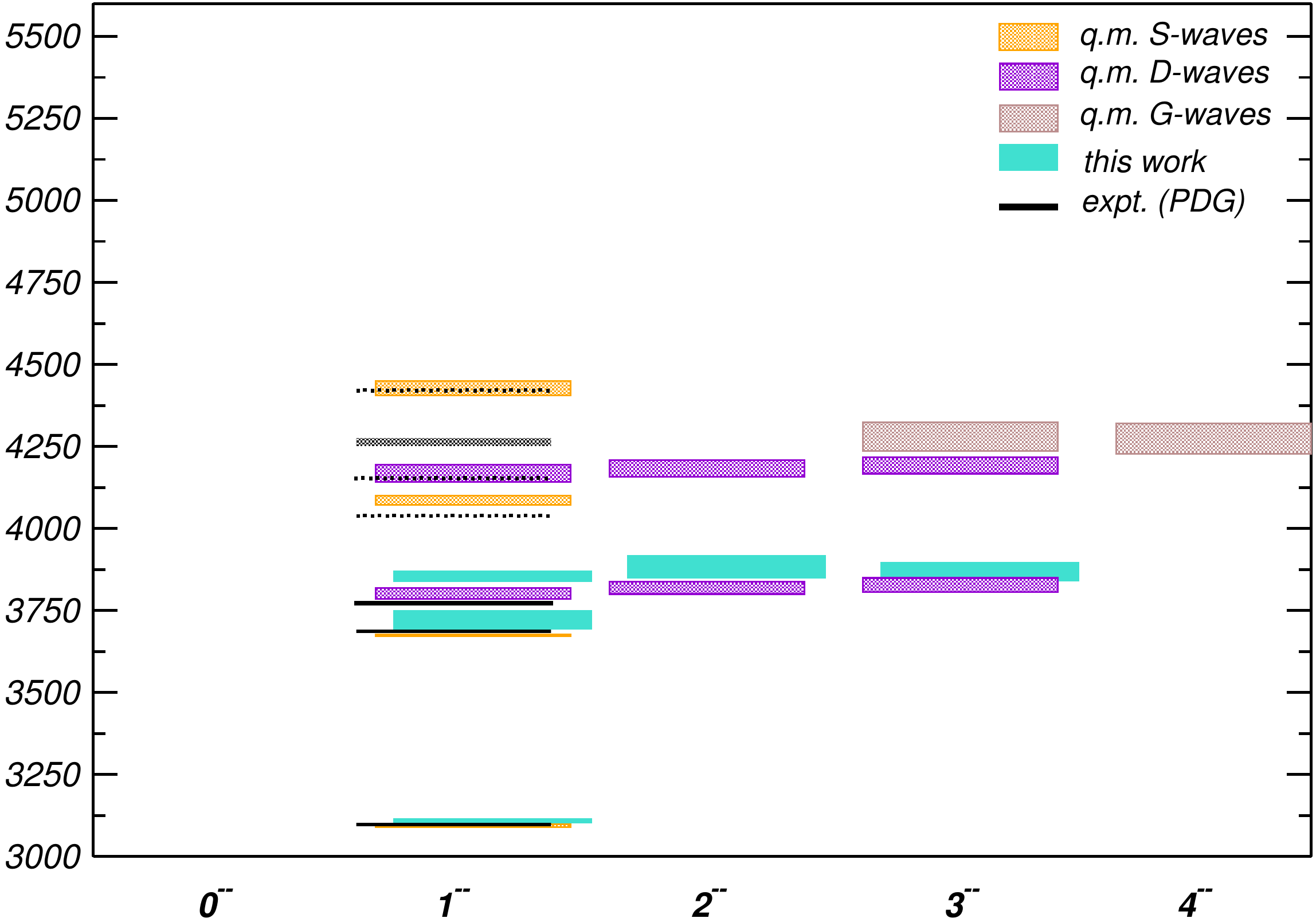}\hfill
    \end{minipage}
&
    \begin{minipage}{0.45\textwidth}
      \includegraphics[width=\textwidth]{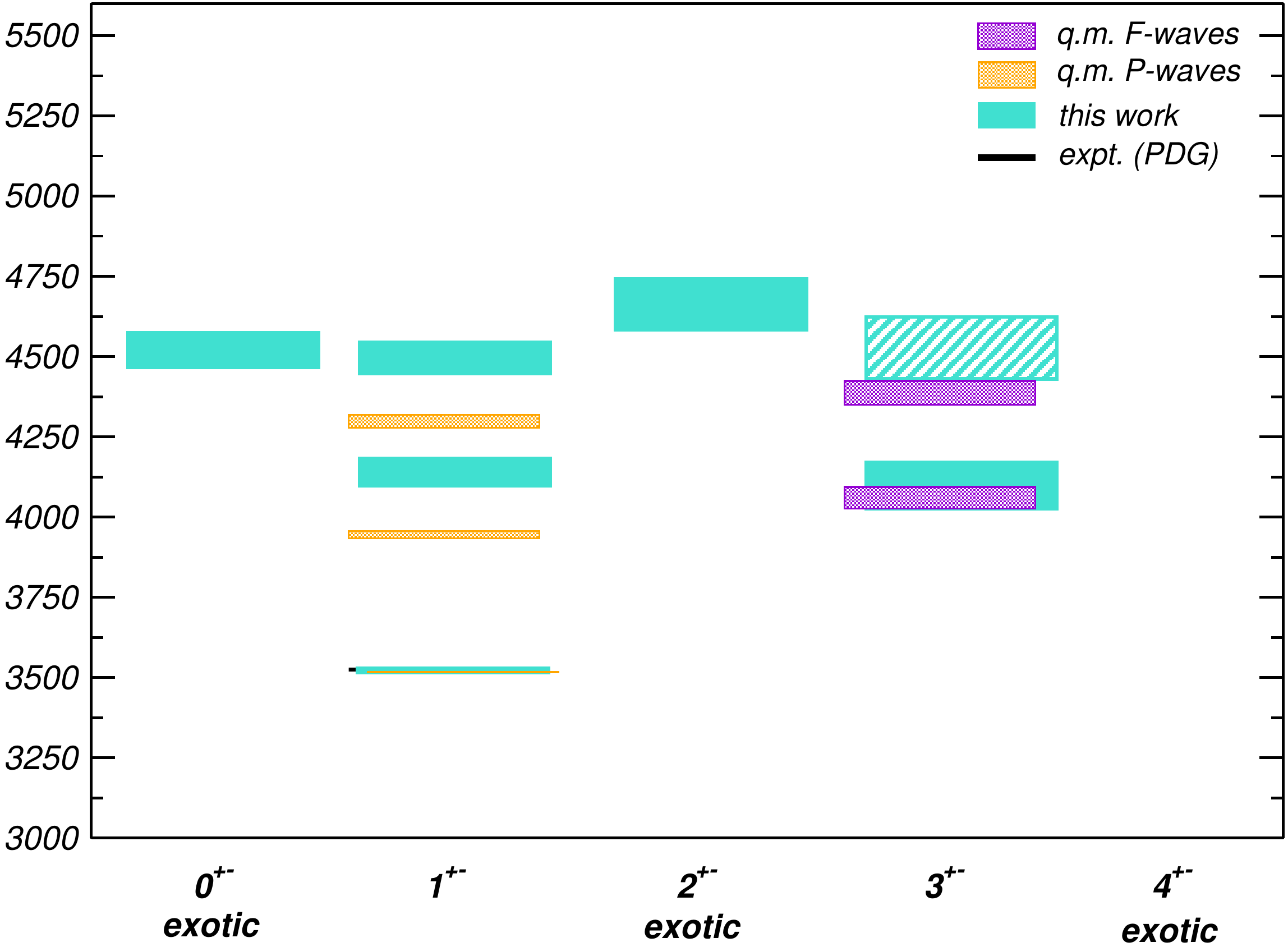}\hfill
    \end{minipage}
  \end{tabular}
  \caption{Charmonium excitation spectrum in MeV from
    Ref.~\protect\refcite{Dudek:2007wv}.  Left: $J^{--}$ states.
    Right $J^{+-}$ states including exotics.  Lattice results (blue)
    are compared with PDG values (black) and various quark model
    results (purple and orange). Color online.
    \label{fig:Dudek}
  }
\end{figure}

\section{Excitation spectrum and exotics}

In 2008 Dudek \et\cite{Dudek:2007wv} published results of a
comprehensive lattice QCD study of the excitation spectrum of
charmonium.  In order to access excited states they introduced a large
set of interpolating operators and used a ``variational'' method to
determine their masses.

How well does this work?  Figure~\ref{fig:Dudek} (left) gives an
impression.  Consider the $1^{--}$ channel, where the ground state
$J/\psi$ has a very clean signal.  Six states were found, but only the
first three shown were unambiguous enough that the authors were
willing to associate them with known levels, namely the $J/\psi$,
$\psi(2S)$ and $\psi(3S)$.  The lattice levels were generally higher
than experiment and the discrepancies grew from 12 to 82 MeV.  They
were also higher than quark model determinations, as shown.  In the
$PC = +-$ channel this study also produced a couple spin-exotic states
as shown in Fig.~\ref{fig:Dudek}.  See Ref.~\refcite{Dudek:2007wv} for
more states.

These results are pioneering, but they are deficient for a number of
reasons.  Sea quark effects were omitted, the calculation was done at
only one lattice spacing, so an extrapolation to the continuum is not
possible, and open charm states were ignored.  The Hadron Spectrum
Collaboration is remedying these shortcomings\cite{Ryan}.

To include sea quarks and allow extrapolation to the continuum and to
physical light quark masses requires generating gauge field
configurations (ensembles) with various lattice spacings and light
quark mass combinations.  The MILC collaboration makes publicly
available a large archive of such ensembles with lattice spacing
ranging from 0.15 fm to 0.045 fm and light quark mass ratios
$m_{ud}/m_s$ ranging down to 0.05, compared with the physical value of
approximately 0.037\cite{Bazavov:2009bb}.

MILC gauge configurations are being used by the FNAL/MILC and HPQCD
collaborations to study the quarkonium spectrum.  The FNAL/MILC
campaign currently uses Fermilab quarks for both charm and bottom and
HPQCD uses HISQ charm quarks and NRQCD bottom quarks.  Most of the
FNAL/MILC results presented here are from last year's
study\cite{Burch:2009az}, which, like the HPQCD study, used a limited
set of interpolating operators and looked at only the low-lying
states.  A new FNAL/MILC campaign currently under way is aimed at
excited as well as ground states using a large set of interpolating
operators, higher statistics, smaller lattice spacings, and more
accurate heavy quark mass tuning.  I will give a preview of the
improvement we hope to obtain.

\begin{figure}
  \centering
  \begin{tabular}{cc}
    \begin{minipage}{0.45\textwidth}
  \includegraphics[width=\textwidth]{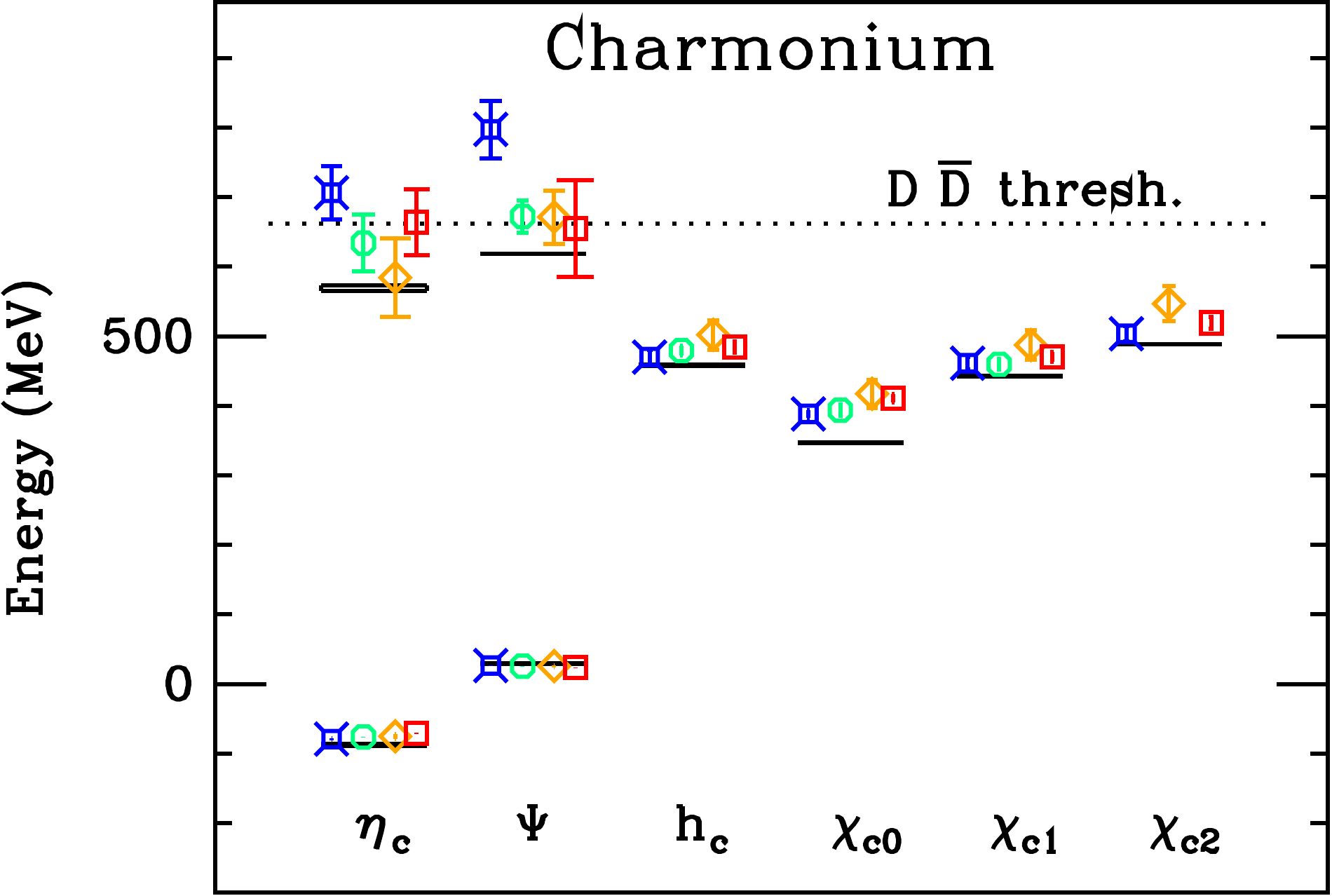}\hfill
    \end{minipage}
&
    \begin{minipage}{0.48\textwidth}
  \includegraphics[width=\textwidth]{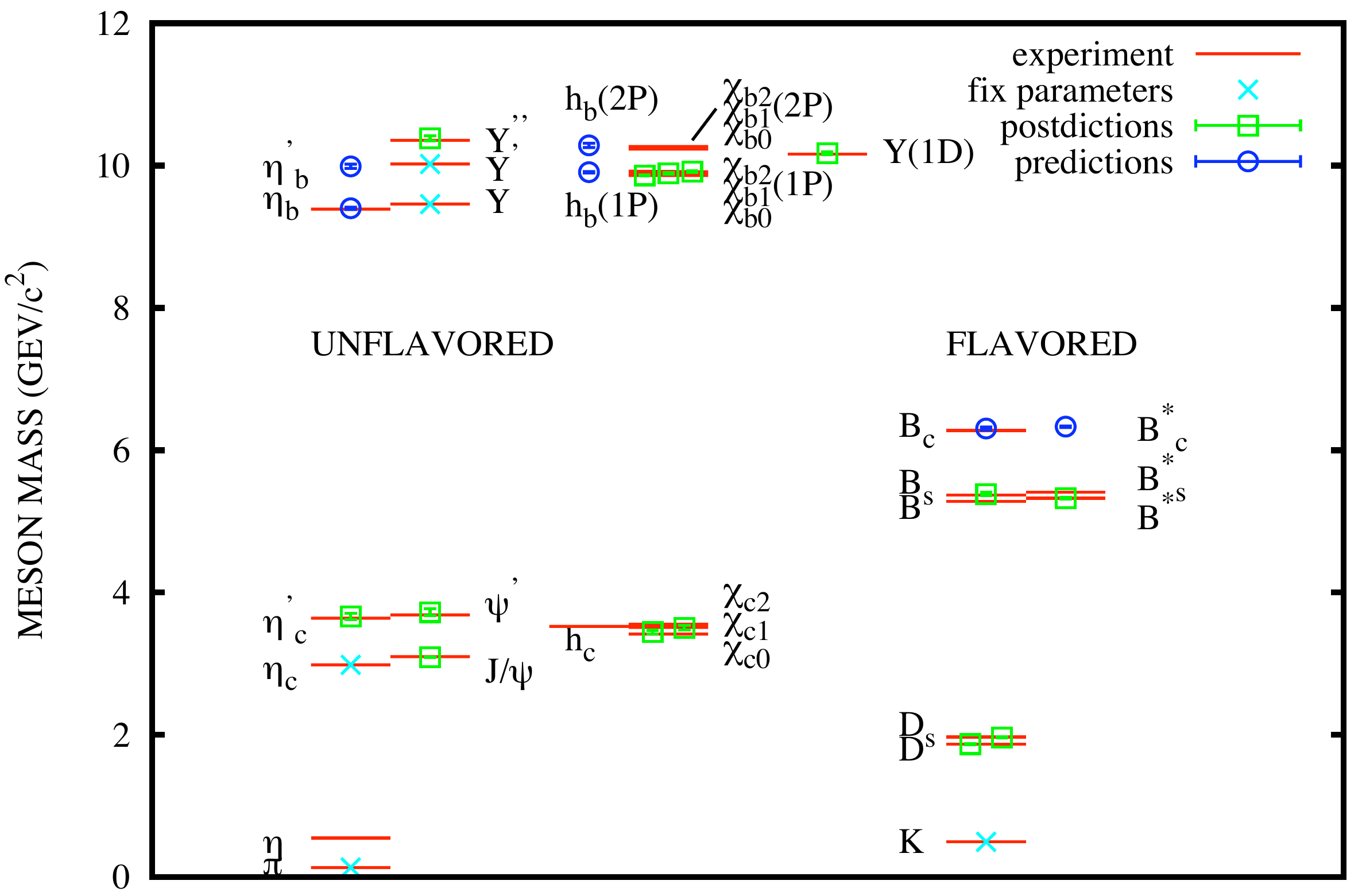}
    \end{minipage}
  \end{tabular}
  \caption{Left: Charmonium levels from
    Ref.~\protect\refcite{Burch:2009az} based on splittings from the
    spin-averaged $1S$ level with quark masses determined from the
    $D_s$ and light pseudoscalar mesons and scale determined from
    upsilon splittings.  Results for four lattice spacings from 0.18
    fm (red) to 0.09 fm (blue) are shown.  Black lines are PDG values.
    Right: complete low-lying heavy and light meson spectrum from
    Ref.~\protect\refcite{Gregory:2009hq} showing five masses used to
    fix four quark masses and the lattice scale and three masses
    predicted in advance of experiment.
    \label{fig:overview}}
\end{figure}

\begin{figure}
  \vspace*{-7mm}
  \centering
  \includegraphics[width=0.45\textwidth]{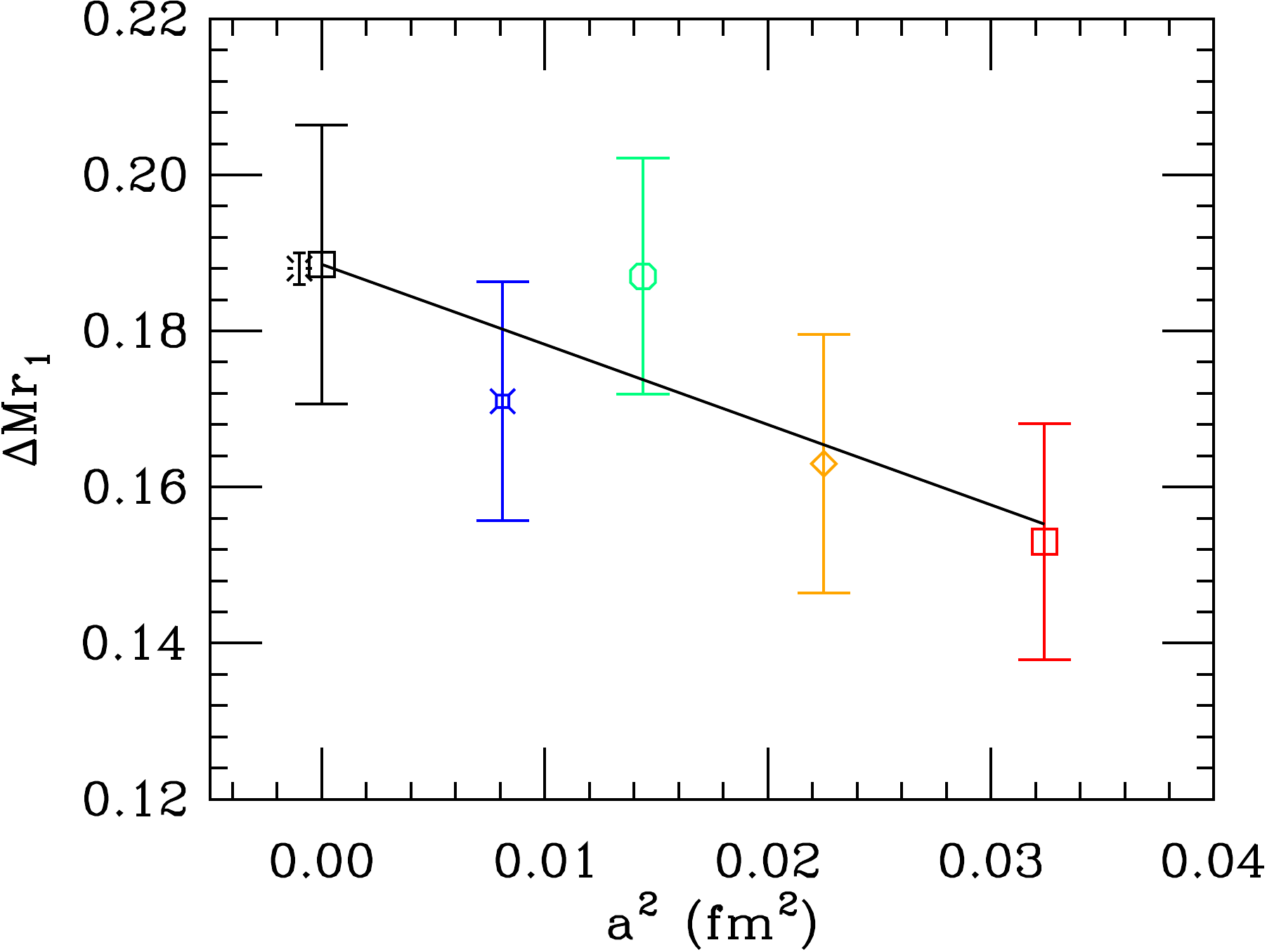}\hfill
  \includegraphics[width=0.45\textwidth]{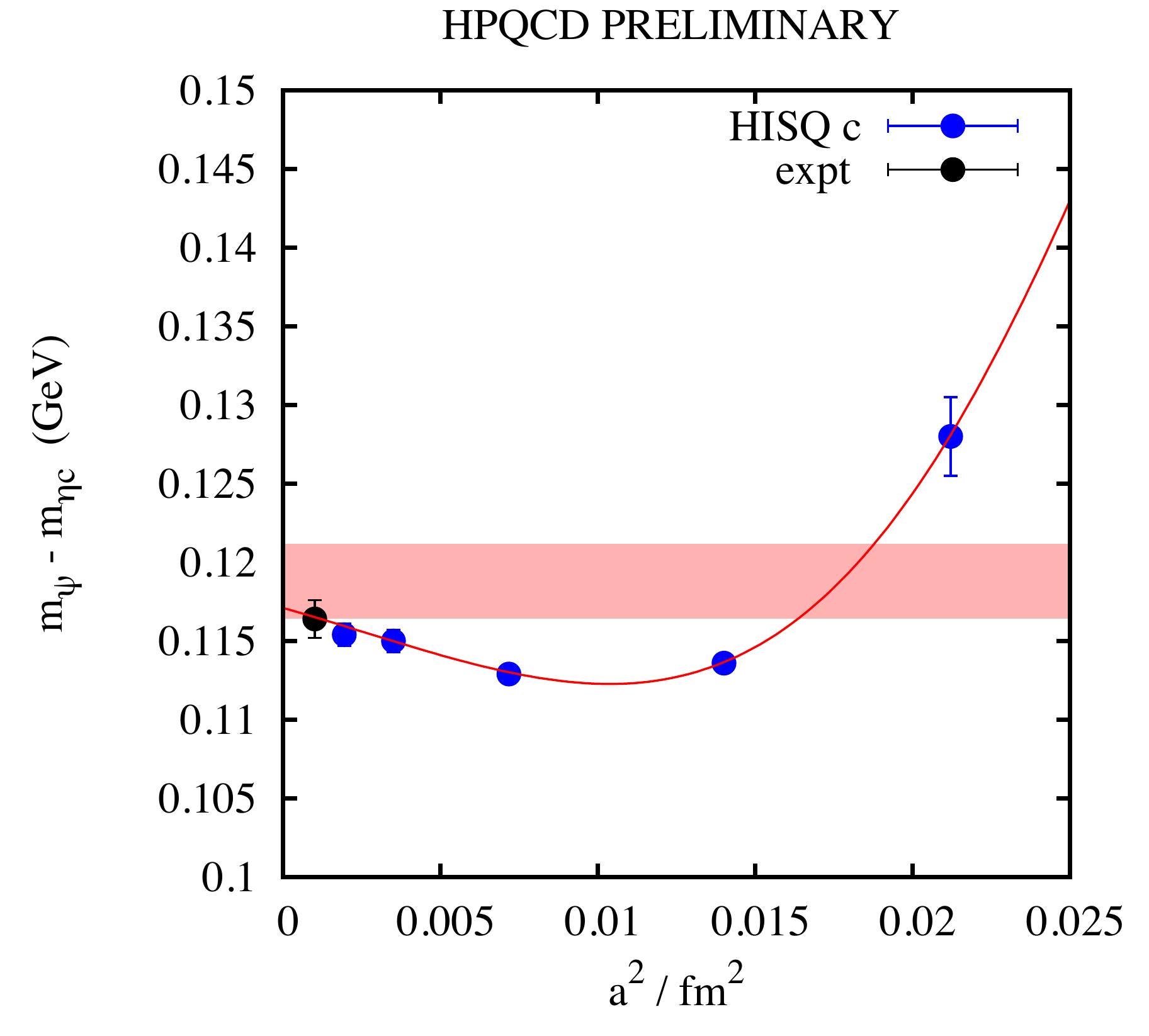}\hfill
  \caption{Hyperfine splitting in charmonium extrapolated to zero
    lattice spacing.  PDG values are in black. Left:
    FNAL/MILC\protect\cite{Burch:2009az}.  Units are $r1 \approx 1.58$
    Gev$^{-1}$.  Right: HPQCD\protect\cite{Follana}.
   \label{fig:HFS}}
\end{figure}

\section{Low-lying levels}

Broadly speaking, lattice simulations do quite well reproducing the
low-lying hadronic states.  Figure~\ref{fig:overview} gives an overview
of the charmonium spectrum of the FNAL/MILC collaboration (left)
\cite{Burch:2009az} and the full gold-plated meson spectrum of the
HPQCD collaboration\cite{Gregory:2009hq} (right).  To appreciate the
precision of the calculation, we turn to a finer mass scale in the
remainder of this minireview.

Historically, the hyperfine splitting of the $1S$ state has proven to
be a stringent test of lattice methodology.  In Fig.~\ref{fig:HFS}
recent results of the FNAL/MILC collaboration (left)
\cite{Burch:2009az} are compared with still more recent results of the
HPQCD collaboration (right)\cite{Follana}.  The FNAL/MILC calculation
was based on Fermilab quarks. The extrapolated value is 117(11) MeV.
In this result annihilation effects are not included.  The HPQCD
result is based on HISQ charm quarks.  The pink band indicates the
full error budget ($\pm 2$ MeV), and includes a small increase due to
annihilation, as suggested by perturbation theory.  Determining level
shifts due to annihilation is difficult.  A recent nonperturbative
(lattice) result estimates, instead, a decrease of $1-4$ MeV from
charm annihilation\cite{Levkova:2010ft}.  Thus, apparently, we have
reached the point where annihilation effects cannot be ignored.

\begin{figure}
  \includegraphics[width=0.42\textwidth]{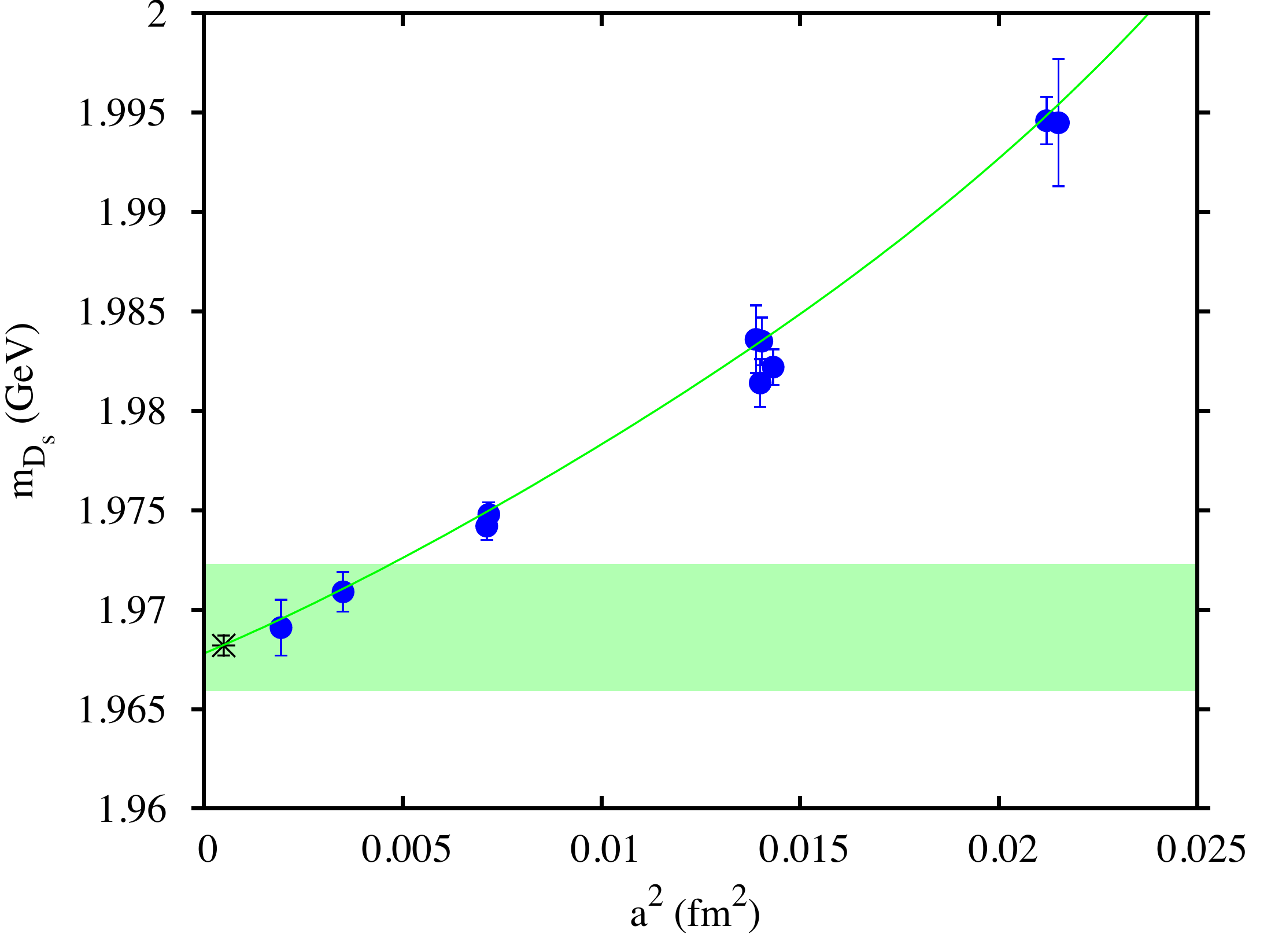}\hfill
  \includegraphics[width=0.42\textwidth]{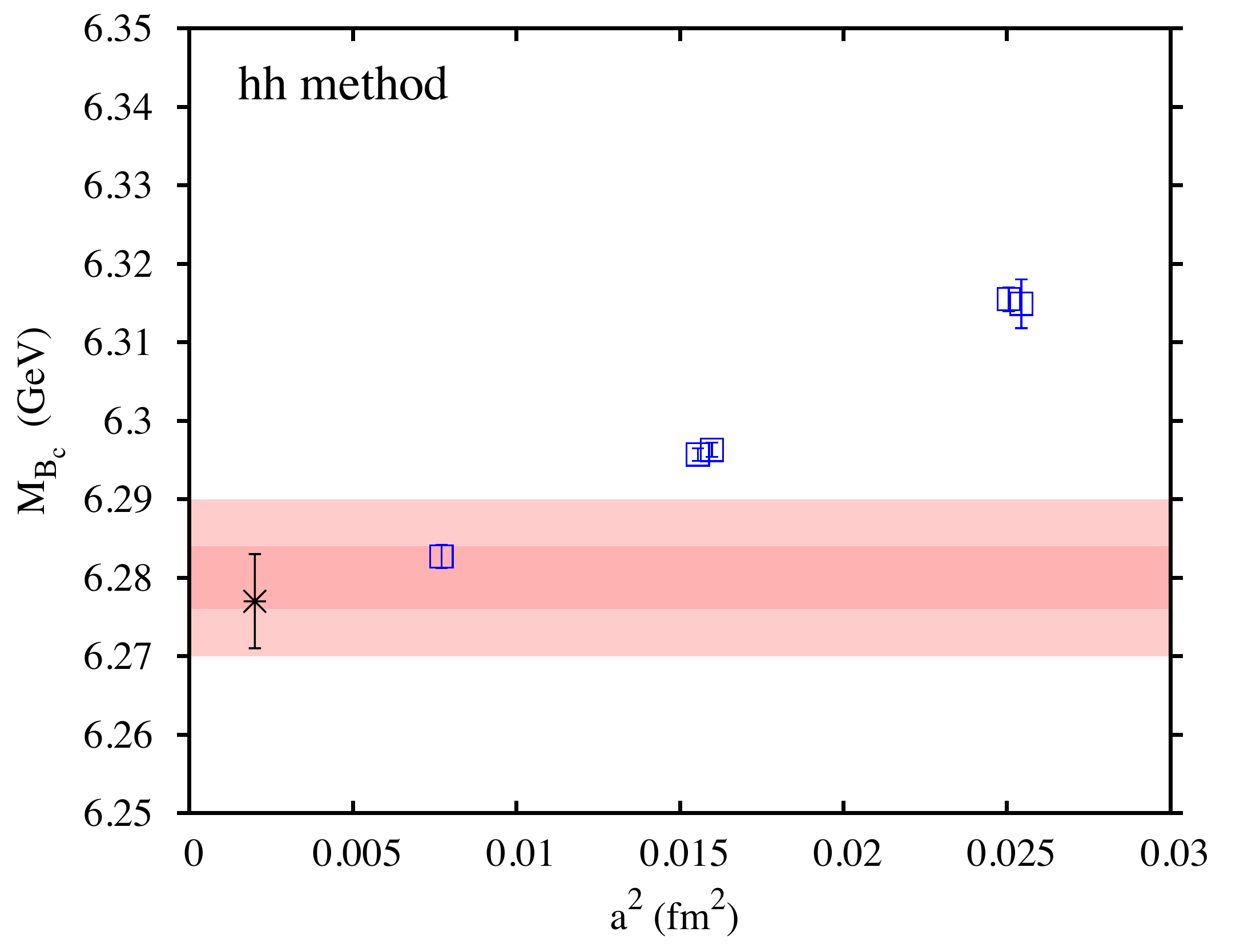}\hfill
  \caption{Recent HPQCD results for the $D_s$ mass
    (left)\protect\cite{Davies:2010ip} and $B_c$
    (right)\protect\cite{Gregory:2010gm}.
   \label{fig:DsBc}}
\end{figure}

Further examples of achievable precision in gold-plated quantities are
shown in Fig.~\ref{fig:DsBc}.  The $D_s$ mass is obtained to 3 MeV
accuracy from the mass splitting $M(c\bar s) -
\frac{1}{2}M(\eta_c)$\cite{Davies:2010ip}, and the $B_c$ mass is
obtained to 10 MeV accuracy from the splitting $M(B_c) - M(b\bar b)/2
- M(\eta_c)/2$\cite{Gregory:2010gm}.

\begin{figure}
  \vspace*{-5mm}
  \includegraphics[width=0.45\textwidth]{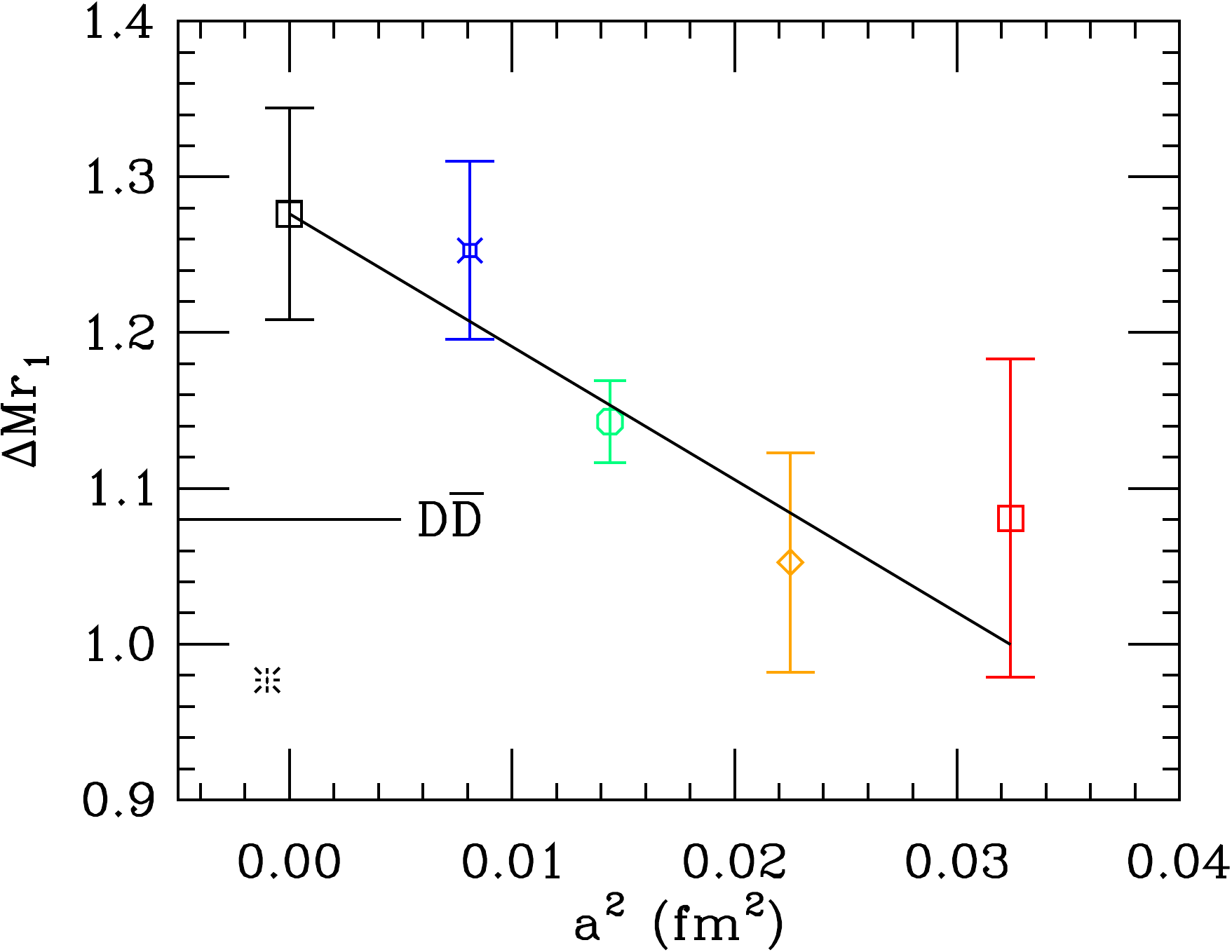}\hfill
  \includegraphics[width=0.45\textwidth]{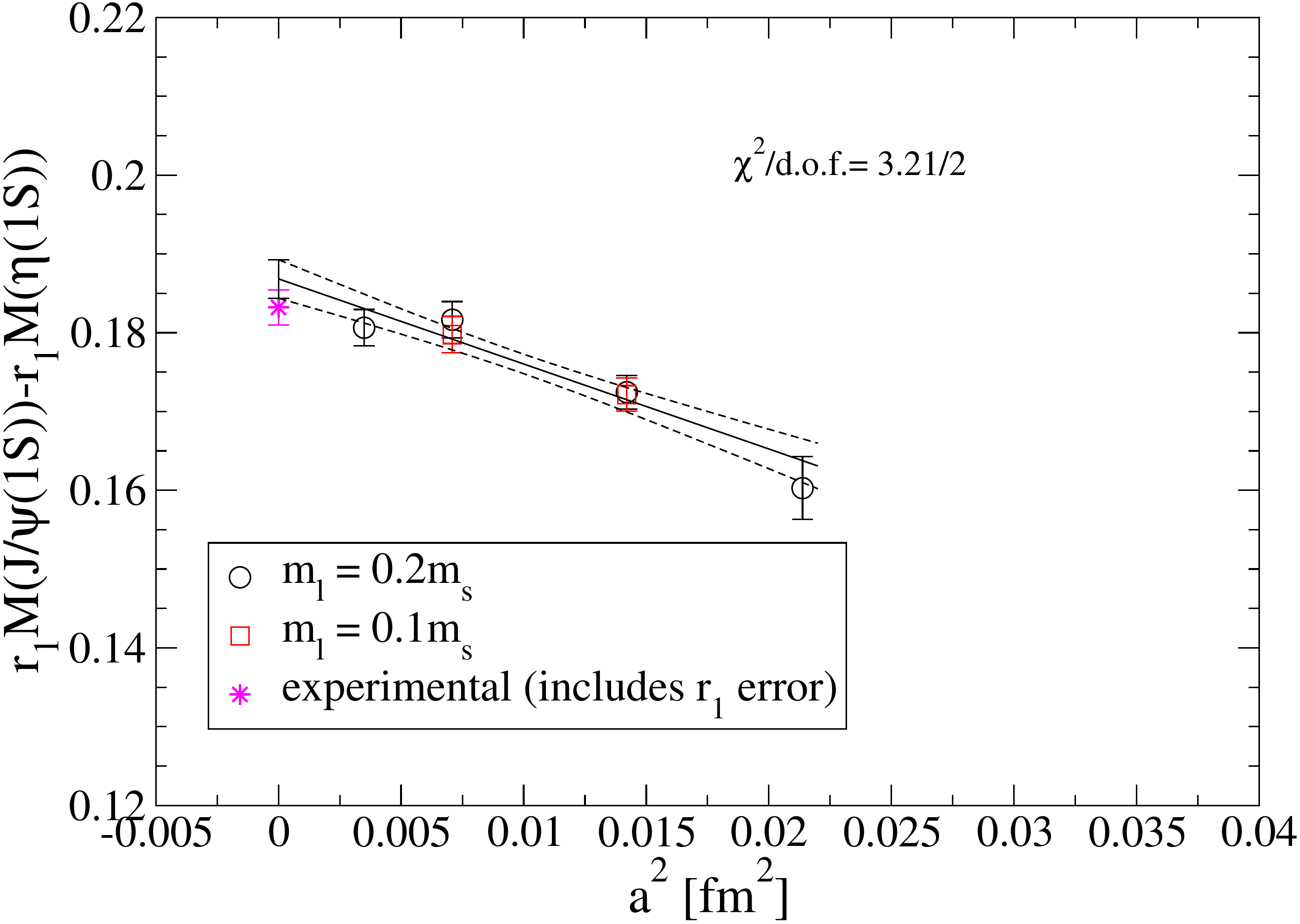}\hfill
    \caption{Left: $\bm{\overline{2S}-\overline{1S}}$ splitting from
      Ref.~\protect\refcite{Burch:2009az} showing possible complications from
      open charm. Right: Sample FNAL/MILC preliminary result for the $1S$
      hyperfine splitting from the new analysis with much reduced errors.
   \label{fig:2S1SnewHFS}}
  \end{figure}

Going beyond gold-plated quantities, the FNAL/MILC simulation results
for the spin-averaged $2S-1S$ splitting, shown in
Fig.~\ref{fig:2S1SnewHFS} (left) disagrees with the experimental
value\cite{Burch:2009az}. Could this be a result of complications from
the nearby open charm threshold?  Those simulations do not currently
treat two-body scattering states.  A recent calculation by Bali and
Ehmann with a small basis set suggests that they may be important for
even states well below threshold, but further study is
needed\cite{Bali:2009er}.

\section{Conclusions and Outlook}

We have seen dramatic improvement in our ability to reproduce the
masses of the lowest lying charmonium states.  This has been brought
about by the use of improved quark actions, higher statistics,
accurate tuning of the heavy quark masses, smaller lattice spacing,
and the availability of gauge field ensembles that support an
extrapolation to physical sea quark masses and zero lattice spacing.
New analysis campaigns with multiple interpolating operators are
underway both by the Hadron Spectrum Collaboration and the Fermilab
Lattice/MILC collaborations.  A preview of the hyperfine splitting
from this new analysis is shown in Fig.~\ref{fig:2S1SnewHFS} (right), a
considerable improvement over Fig.~\ref{fig:HFS} (left).  These campaigns
should provide more accurate information about excited and exotic
states.  Treating two-body scattering states will continue to be a
long-term challenge.

\end{document}